\documentclass[11pt]{article}
\usepackage{vietnam}

\providecommand\prd{Phys.~Rev.~D} %% ``D'' is part of the journal name
\providecommand\jcap{JCAP} 
\providecommand\aap{A\&A}            % {A. \& A.} 
\providecommand\apj{ApJ}                 % {Ap. J.}

\def\be{\begin{equation}}
\def\ee{\end{equation}}
\def\bea{\begin{eqnarray}}
\def\eea{\end{eqnarray}}

\newcommand{\CD}{{\cal D}}
\newcommand{\CE}{{\cal E}}

\newcommand{\CQ}{{\cal Q}}

\newcommand{\CM}{{\cal M}}

\newcommand\Heff{H^{\mathrm{eff}}}
\newcommand\Hpeculiarcomov{H_{\mathrm{pec}}^{\mathrm{com}}}
\newcommand\fvir{f_{\mathrm{vir}}}
\newcommand{\average}[1]{\left\langle #1 \right\rangle_{\cal D}}

\newcommand{\initial}[1]{{#1_{\rm \bf i}}}
\newcommand{\rmd}{{\rm d}}

\newcommand\prerefereechangesI[1]{#1}

\newcommand\prerefereechangesII[1]{#1}    

\newcommand{\Photo}{}

\begin{document}
\vspace*{4cm}
\title{Virialization-induced curvature \protect\prerefereechangesI{versus} dark energy}

\author{ \prerefereechangesI{Jan J. Ostrowski${}^{1,2}\ $\footnote{Long-term visit to CRAL}, 
    Boudewijn F. Roukema$^{1,2}\ $\footnote{Short-term visit to CRAL}, 
    Thomas Buchert$^2$} }

\address{\protect\prerefereechangesI{${}^1$Toru\'n Centre for Astronomy, 
  Faculty of Physics, Astronomy and Informatics,
  Nicolaus
  Copernicus University, ul. Gagarina 11, 87-100 Toru\'n, Poland\\
${}^2$Universit\'e de Lyon, Observatoire de Lyon,
Centre de Recherche Astrophysique de Lyon (CRAL), CNRS UMR 5574: Universit\'e Lyon~1 and \'Ecole Normale Sup\'erieure de Lyon, 
9 avenue Charles Andr\'e, F--69230 Saint-Genis-Laval, France}}

\maketitle
\abstracts{
 The concordance model is successful in explaining numerous observable
 phenomena at the price of introducing an \prerefereechangesI{exotic} source of unknown
 origin: \prerefereechangesI{dark} energy. \prerefereechangesI{Dark} energy dominance occurs at recent
 epochs, when we expect \prerefereechangesI{most cosmological} structures to \prerefereechangesI{have} 
 \prerefereechangesI{already formed,
 and thus, when} the \prerefereechangesI{error induced by} 
 forcing the homogeneous 
 \prerefereechangesI{Friedmann--Lema\^{\i}tre--Robertson--Walker (FLRW)} 
 metric \prerefereechangesI{onto}
 the data is expected to be \prerefereechangesI{the} most significant.
 We propose a way to quantify the impact of deviations from homogeneity
 on the evolution of cosmological parameters. \prerefereechangesI{Using}
 a multi-scale partitioning approach and the virialization fraction
 \prerefereechangesI{estimated from} numerical simulations in an \prerefereechangesI{Einstein--de~Sitter}
 model, we obtain an observationally realistic distance modulus over
 redshifts $0<z<3$ \prerefereechangesI{by} a relativistic correction of the FLRW metric.
 }

\section{Introduction}

\prerefereechangesI{The weakness of the anisotropy observed in the cosmic
microwave background (CMB), i.e. the weakness in deviations from 
angular homogeneity, is} widely accepted as 
\prerefereechangesI{one of the strongest 
  justifications} for using the 
\prerefereechangesI{Friedmann--Lema\^{\i}tre--Robertson--Walker (FLRW)} 
metric to describe the expanding
Universe. 
\prerefereechangesI{However, from first principles,}
the non-commutativity of averaging and time differentiation
\prerefereechangesI{is key} to the unresolved issue of the impact of inhomogeneities on average properties of 
\prerefereechangesI{our Universe, which in reality is strongly inhomogeneous at late epochs}.
Explaining
observations within the standard $\Lambda$CDM cosmology requires a new
type of energy
that \prerefereechangesI{violates the 
weak energy condition}. What is 
\prerefereechangesI{normally seen as a 
``very successful theory''} is in fact putting us in \prerefereechangesI{the} very uncomfortable
position of dealing with \prerefereechangesI{a new type of} energy of unknown 
\prerefereechangesI{origin. There is} no direct
evidence of \prerefereechangesI{the} existence \prerefereechangesI{of what} 
\prerefereechangesI{has recently become} the dominating
component of \prerefereechangesI{the} cosmic \prerefereechangesI{triangle---``crisis''} 
seems to be \prerefereechangesI{a} more adequate word than \prerefereechangesI{``success''}
for the situation in \prerefereechangesI{today's} standard cosmology. Not being subject to
any physical \prerefereechangesI{constraints}, 
\prerefereechangesI{the} cosmological constant comes out as a fitting
parameter rather then \prerefereechangesI{an actual} feature of \prerefereechangesI{the} 
real Universe. We propose a
simpler model (Roukema, Ostrowski \& Buchert 2013 \cite{ROB13})
that matches the observations 
\prerefereechangesI{to a satisfactory initial level of accuracy}
without \prerefereechangesI{the} need for modifying the well-tested \prerefereechangesI{theory of general relativity}.

\newcommand\sectiondescriptionnotneeded{
  The \prerefereechangesI{presentation here, based on Roukema, Ostrowski \& Buchert (2013),}
  is organised as follows: in \prerefereechangesI{Sect.~1} we briefly describe the
  \prerefereechangesI{underlying} theory, in \prerefereechangesI{Sect. 2} we describe the 
  \prerefereechangesI{Virialisation
    Approximation} model, in \prerefereechangesI{Sect. 3} we present the 
  \prerefereechangesI{main}
  result, \prerefereechangesI{and we conclude in Sect. 4}.
}

%\sectiondescriptionnotneeded

\section{Theoretical framework}
\subsection{Scalar averaging}

%\prerefereechangesI{The theory of general relativity} 
%\prerefereechangesI{is based on four-dimensional (4D)} tensors. 
%\prerefereechangesI{For certain}
%\prerefereechangesI{purposes,} it is 
%\prerefereechangesI{useful} to introduce a 3+1
%\prerefereechangesI{foliation, decomposing the} Einstein equations into 
%\prerefereechangesI{spatial}
%constraints and dynamical evolution equations.  One of the advantages
%of this approach is 
%\prerefereechangesI{that,
%once we have chosen} a way to foliate the \prerefereechangesI{4D} spacetime,
%the notion of \prerefereechangesI{a spatial} average can be 
%\prerefereechangesI{uniquely} defined. 
In this section we
will briefly present the scalar averaging procedure 
(see e.g.,
\prerefereechangesI{Buchert 2000} \cite{Buch2000})
treating the scalar \prerefereechangesI{quantities derived
from the} Einstein equations for a dust matter model in a $3+1$ comoving-synchronous slicing of space-time.

We take the spatial average of scalar fields $\Psi$, 
\prerefereechangesII{where}
$ \average{\Psi(t,X^{k})}:=\frac{1}{V_{\cal D}}\int_{\cal D}\rmd\mu_{g}\;\;\Psi(t,X^{k})$,
with the Riemannian volume element of the spatial metric $g_{ij}$, $\rmd \mu_{g}=J \rmd^{3}X$ in local coordinates $X^i$, $J:= \sqrt{\det (g_{ij})}$, and note the non-commutativity relation, $
  \partial_{t}\average{\Psi(t,X^{k})}-\average{\partial_{t}\Psi(t,X^{k})}=\average{\theta\Psi}-\average{\theta}\average{\Psi}$,
where $\theta$ denotes the trace of the expansion tensor. The volume scale factor $a_{\cal D}$, defined via the
domain's volume $V_{\cal D}(t)=|{\cal D}|$, and the initial volume
$V_{\initial{\cal D}}=V_{\cal D}(\initial t)=|\initial{\cal D}|$, $ a_{\cal D}(t):=\left(\frac{V_{\cal D}(t)}{V_{\initial{\cal D}}}\right)^{1/3}$, \prerefereechangesII{obeys} the
well-known equations:
\begin{equation}
  3\frac{{\ddot{a}}_{\cal D}}{a_{\cal D}}+4\pi
  G\frac{M_{\initial{\cal D}}}{V_{\initial{\cal D}}a_{\cal D}^{3}}-\Lambda=\CQ_{\cal D}\;;
\label{eq:expansion-law-GR}
\end{equation} 
\begin{equation}
 3 \left(\frac{{\dot{a}}_{\cal D}}{a_{\cal D}}\right)^{2}+ \frac{3 k_{\initial{\cal D}}}{a_{\cal D}^2} - {8\pi
    G}\frac{M_{\initial{\cal D}}}{V_{\initial{\cal D}}a_{\cal D}^{3}}+\frac{\average{\cal W}}{2}-\Lambda=-\frac{{\cal Q}_{\cal D}}{2}\;,
\label{eq:hamiltonconstraint}
\end{equation}%
where the total rest mass $M_{\initial{\cal D}}$ and the backreaction variables
$\average{\cal W}$ and the ${\cal Q}_{\cal D}$ are \prerefereechangesII{domain-depen\-dent} and, except 
\prerefereechangesII{for} the mass, \prerefereechangesII{are time dependent}.
The backreaction source terms are defined by:
\begin{equation} {\cal Q}_{\cal D}:=\frac{2}{3}\average{\left(\theta-\average{\theta}\right)^{2}}-2\average{\sigma^{2}}\quad;\quad W_{\cal D} =  \average{\cal R} - \frac{6 k_{\initial{\CD}}}{a_{\cal D}^2} \;\;,
\label{eq:Q-GR}
\end{equation}
with $\average{\cal R}$ the averaged spatial 3-Ricci scalar, and $\sigma^2 : = 1/2 \sigma_{ij}\sigma^{ij}$ the squared rate of shear. The backreaction variables \prerefereechangesII{satisfy} the integrability condition:
\begin{equation}
 \partial_{t}{\cal Q}_{\cal D}+6 H_{\cal D} {\cal Q}_{\cal D}+\partial_{t}\average{\cal W}+2 H_{\cal D} \average{\cal W}=0\;\;\;;\;\;\; H_{\cal D} : = \frac{\dot{a}_{\cal D}}{a_{\cal D}}\;\;.
\label{eq:integrability-GR}
\end{equation}

\subsection{Multi-scale partitioning}

\prerefereechangesI{We model the} \prerefereechangesI{large-scale} structure of the Universe 
\prerefereechangesI{by introducing a volume-partitioning into}
two main components,
\prerefereechangesI{defined by dividing a spatial slice into disjoint unions of two complementary
subregions:}
voids and overdense regions. In order to quantify
their behavior, we introduce three characteristic \prerefereechangesI{scales,} 
\prerefereechangesI{$L_{\CE}$} for
voids, 
\prerefereechangesI{$L_{\CM}$} for a typical galaxy cluster and 
\prerefereechangesI{$L_{\CD}$} for 
\prerefereechangesI{the largest scale, at which we 
assume statistical homogeneity on a given spatial hypersurface.}
Defining
\prerefereechangesI{$\lambda_{\CM}$}, the fraction of volume occupied by virialized
matter \prerefereechangesII{as a proportion of}
the total volume, 
\prerefereechangesI{and} using the partitioning rule, $\langle
f \rangle_{\CD} = (1-\lambda_{\CM})\langle f \rangle_{\CE} + \lambda_{\CM} \langle
f \rangle_{\CM}$, we obtain \prerefereechangesI{relations} between averaged
quantities on different domains (Buchert and Carfora 2008 \cite{BuchCar}):
\prerefereechangesI{
\bea
\average{\varrho} = \lambda_{\CM} \langle \varrho \rangle_{\CM} + (1-\lambda_{\CM})\langle \varrho \rangle_{\CE} \;\;,\nonumber\\
\average{\cal R} = \lambda_{\CM} \langle {\cal R} \rangle_{\CM} + (1-\lambda_{\CM})\langle {\cal R} \rangle_{\CE} \;\;, \nonumber\\
H_{\cal D} = \lambda_{\CM} H_{\CM} + (1-\lambda_{\CM}) H_{\CE} \;\;, \nonumber \\
{\cal Q}_{\cal D} = \lambda_{\CM} {\cal{Q}}_{\CM} + (1-\lambda_{\CM}){\cal{Q}}_{\CE} +6\lambda_{\CM}(1-\lambda_{\CM})(H_{\CM} -H_{\CE})^2 \;\;.
\eea
}
  
\section{Virialization Approximation}

The \prerefereechangesI{virialization approximation} (VA) is a hybrid model that 
\prerefereechangesI{uses}
observational, numerical and phenomenological inputs 
\prerefereechangesI{to evaluate the}
underlying analytical model. \prerefereechangesI{This approach is far from}
perfect and needs to be treated as a step \prerefereechangesII{towards} \prerefereechangesI{a} more
refined \prerefereechangesI{model}. In this section we \prerefereechangesI{briefly} list \prerefereechangesI{the}
assumptions used in \prerefereechangesI{the} VA.

\subsection{Observational inputs}
Observational inputs 
\prerefereechangesI{(Table~\ref{tab:obsparams})} are taken from publicly available catalogues and sky surveys, 
\begin{table}[h]
  \caption{Observational \prerefereechangesI{parameters}}
  \label{tab:obsparams}
  \vspace{0.4cm}
  \begin{center}
    \begin{tabular}{lcc}
      \hline
      & parameter & value \\
      \hline
      observational inputs: &low-redshift limit of $\Heff$ \rule{0ex}{2.5ex} % strut
      & $74.0 \pm 1.6$ km/s/Mpc  \\
      &comoving void radius & $25 \pm 2h^{-1}$ Mpc  \\
      &infall vel. around rich cluster & $1200 \pm 30$ km/s \\
      \hline
      inferred value: & zero-redshift value of $\Hpeculiarcomov$ \rule[-1.2ex]{0ex}{3.5ex} % strut
      & $36 \pm 3$ km/s. \\
      \hline
    \end{tabular}
  \end{center}
\end{table}
\prerefereechangesI{where $h := H_0/(100 $~km/s/Mpc$)$ is the dimensionless zero-redshift
  FLRW Hubble parameter. 
  See Roukema, Ostrowski \& Buchert (2013) for details including references.}

\subsection{Analytical assumptions}

We employ \prerefereechangesI{scalar} averaging together with multi-partitioning,
\prerefereechangesI{supplemented by} some reasonable simplifications. We use
the \prerefereechangesI{(mass-based) virialization fraction} 
$\fvir$ \prerefereechangesI{estimated} from 
\prerefereechangesI{Einstein--de~Sitter} (EdS) $N$-body simulations and the
density contrast 
\prerefereechangesI{of non-linear collapse}
$\delta_{vir}\approx 200$ from \prerefereechangesI{the} scalar virial theorem
to determine the $\lambda_M$ parameter. \prerefereechangesI{Adopting} the stable clustering
hypothesis \prerefereechangesI{(Peebles 1980 \cite{Peebles1980})}, 
\prerefereechangesI{i.e. $H_{\CM} \approx 0$,} considering the kinematical
backreaction term ${\cal Q}_{\cal D}$ as subdominant,
\prerefereechangesI{and defining $\lambda_{\CM}: = \frac{\fvir}{\delta_{\mathrm{vir}}}$,}
results in a closed set of balance equations:
\bea
\Omega_{\cal R}^{\CD} = \lambda_{\CM} \Omega^{\CM}_{\cal R} + (1-\lambda_{\CM})\Omega^{\CE}_{\cal R} - \frac{\lambda_{\CM}}{1-\lambda_{\CM}}\;\;,\nonumber \\
 \Omega_{\mathrm m}^{\CD} = \lambda_{\CM} \Omega^{\CM}_{\mathrm m} + (1-\lambda_{\CM})\Omega^{\CE}_{\mathrm m}\;\;,\nonumber \\
\Omega_{\mathrm m}^{\CM} + \Omega_{\cal R}^{\CM} = 0 \;\;,\;\;
\Omega_{\mathrm m}^{\CD} + \Omega_{\cal R}^{\CD} = 1\;\;.
\eea
 
\section{Results}

\begin{figure}  % [ht]
  \centering
  \includegraphics[width=8cm]{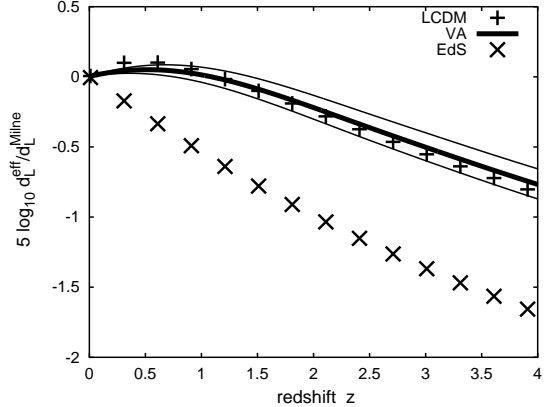} %% latex + dvips + ps2pdf with 
  \caption[]{\protect\prerefereechangesI{Distance modulus as a function of redshift $z$,
      for two FLRW models (i.e. uncorrected for virialization), the
      $\Lambda$CDM model (``$+$'' symbols,
      approximately fitting supernova type Ia observations),
      and the EdS model (``$\times$'' symbols), and for the EdS VA
      (thick curve) with approximate errors (thin curves).}}
  \label{f-fig1}
\end{figure}

The main result obtained with \prerefereechangesI{the} VA is the effective metric
\prerefereechangesI{[(2.37), Roukema, Ostrowski \& Buchert 2013]}. Here \prerefereechangesI{we}
present only \prerefereechangesI{one} possible use of the effective metric, namely the
\prerefereechangesI{redshift--distance-modulus}
\prerefereechangesI{relation---a} standard tool \prerefereechangesI{in}
observational cosmology, \prerefereechangesI{shown in Fig.~\ref{f-fig1}}.

\section{Summary \protect\prerefereechangesI{and prospects}}

It is widely believed that the effect of inhomogeneities on \prerefereechangesI{large-scale}
dynamics \prerefereechangesI{is} negligible and cannot explain \prerefereechangesI{away} dark
energy. Moreover, Newtonian perturbations added on top \prerefereechangesI{of}
uniformly expanding space are commonly assumed to describe the
structure formation process \prerefereechangesI{accurately}.
\prerefereechangesI{It has even been suggested that under certain conditions
  on a class of solutions of the Einstein equations, the contributions
  to the stress-energy tensor would be traceless (as tensors) and obey
  the weak energy condition, thus being unable to mimic dark energy
  with its unusual equation of state (Wald \& Green 2011).}
\prerefereechangesI{However, exact, inhomogeneous, dark-energy--free
  cosmological solutions of the Einstein equations have long been
  shown to match the observed redshift--luminosity-distance relation
  (the Stefani model, Dabrowski \& Hendry 1998 \cite{DabHend98}; 
  the Lema\^{\i}tre--Tolman--Bondi model, e.g., {C{\'e}l{\'e}rier}, {Bolejko}, \&
  {Krasi{\'n}ski} 2010 \cite{CBK10}).}

We present \prerefereechangesI{a somewhat}
different,
\prerefereechangesI{though still general-relativistic and
dark-energy--free,} approach 
\prerefereechangesI{focussing} on scalar
quantities \prerefereechangesI{derived from a} generic metric and trying to determine their
evolution
\prerefereechangesI{(cf. Wiltshire et al. 2012 \cite{Wiltshire12Hflow}).}
The VA model is designed to be a rough working model based
on replacing missing puzzles from \prerefereechangesI{a still developing} 
analytical approach with a combination of observable parameters and
$N$-body simulations estimates. \prerefereechangesI{Since observationally} realistic 
\prerefereechangesI{values result from}
this framework, \prerefereechangesI{there is a strong motivation}
\prerefereechangesII{towards} replacing the \prerefereechangesI{unphysical} dark energy
with realistic GR-based properties. Detailed calculations
involving \prerefereechangesI{relativistic Lagrangian perturbation theory} 
\prerefereechangesI{provide} yet
another argument supporting this case
\prerefereechangesI{(Buchert \& Ostermann 2012 \cite{BuchRZA1}; Buchert, Nayet \& Wiegand 2013 \cite{BuchRZA2}).} 
\prerefereechangesI{Applying the relativistic Zel'dovich
approximation to the} VA will allow us to \prerefereechangesI{strengthen}
this model and provide an important voice in 
\prerefereechangesI{the ongoing debate on the relevance} of \prerefereechangesI{inhomogeneities in the} Universe.

%% Please preserve the style of the headings, text fonts and line
%% spacing to provide a uniform style for the proceedings volume.

% Equations should be centered and numbered consecutively, as in
% Eq.~\ref{eq:murnf}, and the {\em eqnarray} environment may be used to
% split equations into several lines, for example in Eq.~\ref{eq:sp},
% or to align several equations.
%An alternative method is given in Eq.~\ref{eq:spa} for long sets of

%\subsection{Figures}\label{subsec:fig}

%If you wish to `embed' an image or photo in the file, you can use
%the present template as an example. The command 
%\verb^\includegraphics^ can take several options, like
%\verb^draft^ (just for testing the positioning of the figure)
%or \verb^angle^ to rotate a figure by a given angle.

%The caption heading for a figure should be placed below the figure.

\section*{Acknowledgments}

Part of this work consists of research conducted in the scope of the
HECOLS International Associated Laboratory.  Some of JJO's
contributions to this work were supported by the Polish Ministry of
Science and Higher Education under ``Mobilno\'s\'c Plus II edycja''. A
part of this project has made use of Program Oblicze\'n Wielkich
Wyzwa\'n nauki i techniki (POWIEW) computational resources (grant 87)
at the Pozna\'n Supercomputing and Networking Center (PCSS).
This work was conducted within the 
\prerefereechangesII{``Lyon Institute of Origins'' Grant} No. ANR-10-LABX-66.

%\subsection{Final Manuscript}\label{subsec:final}

%All files (.tex, figures and .pdf) should be sent by the {\bf 15th of October 2013}
%by e-mail 
%to \\
%{\bf moriond@in2p3.fr}.\\

%\

%It should be noted that these are physically identical and
%form just one true parametrization.
%\bea
%T & = & Im[V_{11} {V_{12}}^* {V_{21}}^* V_{22}]  \nonumber \\
%&  & + Im[V_{12} {V_{13}}^* {V_{22}}^* V_{23}]   \nonumber \\
%&  & - Im[V_{33} {V_{31}}^* {V_{13}}^* V_{11}].
%\label{eq:sp}
%\eea

%\begin{figure}
%\begin{minipage}{0.33\linewidth}
%\centerline{\includegraphics[width=0.9\linewidth,draft=true]{figexamp}}
%\end{minipage}
%\hfill
%\begin{minipage}{0.32\linewidth}
%\centerline{\includegraphics[width=0.9\linewidth]{figexamp}}
%\end{minipage}
%\hfill
%\begin{minipage}{0.32\linewidth}
%\centerline{\includegraphics[angle=-45,width=0.9\linewidth]{figexamp}}
%\end{minipage}
%\caption{same figure with draft option (left), normal (center) and rotated (right)}
%\label{fig:radish}
%\end{figure}

%\section*{Acknowledgments}

%$\section*{Appendix}

% We can insert an appendix here and place equations so that they are
%given numbers such as Eq.~\ref{eq:app}.
%\be
%x = y.
%\label{eq:app}
%\ee

\end{document}